\documentstyle[aps]{revtex}

\newcommand{\mvk}{\vert\vec {\it k}\vert}

\begin{document}
\draft  

\title{Behavior of logarithmic branch cuts in the self-energy of
  gluons at finite temperature}

\author{F. T. Brandt and J. Frenkel}
\address{Instituto de F\'\i sica,
Universidade de S\~ao Paulo\\
S\~ao Paulo, SP 05315-970, BRAZIL}

\date{\today}
\maketitle 

\begin{abstract}
We give a simple argument for the cancellation of the
$\log(-k^2)$ terms ($k$ is the gluon momentum) between the 
zero-temperature and the temperature-dependent parts of the
thermal self-energy.
\end{abstract}

\bigskip
\noindent
There have been many studies of thermal Green functions in gauge
field theories
\cite{weldon:1982aq,braaten:1990azbraaten:1992gm,frenkel:1990brfrenkel:1991ts,baier:1992dykobes:1992ys,blaizot:1997az,kapusta:book89,lebellac:book96},
which show that their behavior at finite temperature is rather
different from the one at zero temperature. In particular, it was
recently pointed out by Weldon \cite{weldon:1999bs} that in QED, the
logarithmic branch cut singularities cancel to one loop-order, in the thermal
self-energy of the electron.

The purpose of this note is to show that in the Yang-Mills theory, a somewhat
similar behavior occurs in the full gluon self-energy, which
includes  finite temperature effects. Of course, in this theory, the 
massless gluons are quite modified by these effects and the gluon propagator
requires the Braaten-Pisarski resummation. Nevertheless, it is
interesting to remark that, even before such a procedure is carried out,
the one-loop $\log(-k^2)$ terms cancel in the sum of the $T=0$ and the 
$T\neq 0$ contributions to the gluon self-energy. As we shall see, this happens
because the $\log(-k^2)$ terms appear in the thermal part of the
self-energy only in the combination $\log(-k^2/T^2)$. But one can show
that the $\log(T^2)$
contributions have the same structure as the ultraviolet divergent
terms which occur at zero temperature
\cite{brandt:1997rz}. Consequently, the $\log(-k^2/T^2)$ terms combine
directly with the  $\log(-k^2/\mu^2)$ contributions 
which occur at $T=0$ ($\mu$ is the renormalization 
scale), so that the $\log(-k^2)$ terms
cancel in a simple way in the thermal self-energy of the gluon.
The branch cut in the $\log(-k^2)$ contribution at $T=0$ is
associated with the imaginary part of the self-energy, which gives the
rate of decay of a time-like virtual gluon into two real
gluons. Although this contribution cancels at $T\neq 0$, there appear
then additional, temperature-dependent logarithmic branch points. These
singularities indicate processes not available at zero temperature,
where particles decay or are created through scattering in the thermal bath.

To {\it one-loop} order,  the thermal self-energy of gluons generally 
depends on three structure functions, $\Pi^T$, $\Pi^L$ and $\Pi^C$
\cite{weldon:1996kb}
\begin{equation}
  \label{eq:1}
\Pi_{\mu\nu}^{ab}(k_0,\vec k)=  g^2 C_G \delta^{ab}\left(
\Pi^T P_{\mu\nu}^T + \Pi^L P_{\mu\nu}^L 
+ \Pi^C P_{\mu\nu}^C\right),
\end{equation}
where the projection operators $P_{\mu\nu}^{T,L}$
are transverse with respect to the external four-momentum
$k^\mu$ and satisfy: $k^i P_{i\nu}^{T}=0$ and 
$k^i P_{i\nu}^{L}\neq 0$ \cite{kapusta:book89,lebellac:book96}.
The non-trnsverse projection operator
$P^C_{\mu\nu}$ can be written in the plasma rest frame as follows
\cite{weldon:1996kb}
\begin{equation}
  \label{eq:2}
P_{\mu\nu}^C=\frac{1}{k^2}\left[
\frac{k_\nu}{\mvk}\left(k_0k_\mu -\eta_{\mu 0} k^2\right)
+ \mu \leftrightarrow \nu
\right]  .
\end{equation}
Although $\Pi^C$ vanishes at $T=0$ because of the 
Slavnov-Taylor identity, it is in general a non-vanishing function
of the temperature, so that $k^\mu \Pi_{\mu\nu}\neq0$ for the
exact self-energy.

We will discuss here, for definiteness, the retarded thermal self-energy
of the gluon, which is obtained by the analytic continuation 
$k_0\rightarrow k_0+i\epsilon$.(A rather similar analysis can be made
in the case of the time-ordered self-energy, following the approach presented
in reference \cite{dealmeida:1992iu}).
In order to illustrate in a simple way the mechanism of the
cancellation of the $\log(-k^2)$ contributions, let us first consider
the special case of the {\it Feynman gauge}, where $\Pi^C$
vanishes even at finite temperature.
Then, $\Pi^T$ and $\Pi^L$ can be expressed in the plasma rest
frame in terms of linear combinations of 
$\Pi^\mu_\mu$ and $\Pi_{00}$.
After performing the integration over the internal energies
$q_0$, $\Pi^\mu_\mu$ and $\Pi_{00}$ can be written as an 
integral over internal on-shell momenta $q=(|\vec q|,\vec q)$,
as follows 
\begin{equation}
  \label{eq:3}
  \Pi^{\mu\;ab}_\mu = g^2 C_G \delta^{ab}\left(
\frac{T^2}{3} - 10 k^2 I_0
\right)
\end{equation}
and
\begin{equation}
  \label{eq:4}
\Pi_{00}^{ab}=2 g^2 C_G \delta^{ab}|\vec k|^2
\left(I_0+4\,I_1\right).
\end{equation}
where ($x$ is the cosine of the angle between $\vec k$ and $\vec q$).
\begin{equation}
  \label{eq:5}
I_{0,1}=\frac{\mu^\epsilon}{(2\pi)^{3-\epsilon}}
\int\frac{d^{3-\epsilon}\vec q}{2|\vec q|}
\left(\frac{1}{k^2+2q\cdot k} + \frac{1}{k^2-2q\cdot k}\right)
\left[\frac{\vec q^2}{k^2}\left(1-x^2\right)\right]^{0,1}
\left[\frac 1 2 + N\left(\frac{|\vec q|}{T}\right)\right].
\end{equation}
The two terms in the last square bracket are associated
respectively with the $T=0$ and the $T\neq 0$ contributions 
($N$ is the Bose-Einstein distribution).

In order to express the integrations in (\ref{eq:5}) in terms 
of known functions, it is convenient to define the variable
\begin{equation}
  \label{eq:6}
K(x) = \frac{1}{4\pi i}\frac{k^2}{k_0-|\vec k|x}.
\end{equation}
Then  it is straightforward to show that
\begin{equation}
 \label{eq:7}
I_0=\frac{i\pi}{|\vec k|}\left(\frac{1}{4\pi}\right)^{\frac{3-\epsilon}{2}}
\frac{\mu^\epsilon}{\Gamma\left(\frac{3-\epsilon}{2}\right)}
\int_{K_{-}}^{K_{+}} dK \int_0^\infty d |\vec q| 
\frac{|\vec q|^{1-\epsilon}}{|\vec q|^2+(2\pi K)^2}
\left[\frac 1 2 +N\left(\frac{|\vec q|}{T}\right)\right],
\end{equation}
where 
\begin{equation}
  \label{eq:8}
K_{\pm}\equiv  K(\pm 1) = \frac{k_0\pm|\vec k|}{4\pi i}.
\end{equation}
The above form shows that the integrals appearing in the calculation of
the gluon self-energy can be naturally expressed in terms of the
quantities $K_\pm$ (which are proportional to the
light-cone momenta $k_0\pm k_3$, if one chooses, for example, the third axis
along $\vec k$).

The $|\vec q|$ integration of $T=0$ part of $I_0$, gives
\begin{equation}
  \label{eq:9}
\frac{i}{8\pi \mvk}
\int_{K_{-}}^{K_{+}} dK
\left[
  \frac{1}{\epsilon} - \log{\frac{2\sqrt{\pi}K}{\mu}} -
\frac{\gamma}{2}+1
\right].
\end{equation}
Using the fact that ${\rm Re}K(x)>0$, the $|\vec q|$ integration of
the $T\neq 0$ part of $I_0$ (where we may set $\epsilon=0$), 
yields the result \cite{gradshteyn}
\begin{equation}
  \label{eq:10}
\displaystyle{\frac{i}{8\pi \mvk}
\int_{K_{-}}^{K_{+}} dK
\left[ \frac{T}{2K} + \log{\frac K T}
- T \frac{d}{dK}\log{\Gamma\left(1+\frac K T\right)}\right]} ,
\end{equation} 
where the logarithm of the gamma function is analytic when
$K\rightarrow 0$. Then, the approximation  
\begin{equation}
  \label{eq:11}
  N\left(\frac{|\vec q|}{T}\right) = 
\frac{1}{\exp(|\vec q|/T)-1} \simeq 
 \theta(T-|\vec q|)\left(\frac{T}{|\vec q|} - \frac 1 2\right),
\end{equation}
would simply lead, after performing the $|\vec q|$
integration in equation (\ref{eq:7}), to the first two
terms in the exact expression  (\ref{eq:10}).
As far as the $\log(K)$ contribution to the $T\neq 0$ part is
concerned, one may
effectively replace in Eq. (\ref{eq:7}), for small $|\vec q|$, 
$N(|\vec q|/T)$ by $-1/2$. Consequently, this contribution
will cancel the $\log(K)$ term associated with the $T=0$ part of
$I_0$ (this cancellation can also be explicitly verified from Eqs.
(\ref{eq:9}) and (\ref{eq:10})).

By itself, the $K$-integration of the $\log(K/T)$ 
term in Eq. (\ref{eq:10}) gives the contribution
\begin{eqnarray}
  \label{eq:12}
&\displaystyle{\frac{i}{16\pi  \mvk}
\left[\left(K_+ + K_-\right)\log{\frac{K_+}{K_-}}+
\left(K_+ - K_-\right)\log{\frac{K_+K_-}{T^2}} 
-2\left(K_+ - K_-\right)\right]}
=\nonumber\\
&\displaystyle{\frac{1}{32\pi^2}\left[
\frac{k_0}{\mvk}\log{\frac{k_0+\mvk}{k_0-\mvk}} + 
\log{\frac{-k^2}{16\pi^2 T^2}-2}\right]} .
\end{eqnarray}
The emergence of the $\log(-k^2)$ term in the special combination 
$\log(-k^2/T^2)$, is a direct consequence of the fact that the integrand 
in Eq. (\ref{eq:10}) depends only on the
dimensionless ratio $K/T$. Similarly, the $\log(K/\mu)$ term in
Eq. (\ref{eq:9}) yields a contribution which, apart from sign, can be
obtained from Eq. (\ref{eq:12}) by the replacement $T\rightarrow\mu$. 
Consequently, the $\log(-k^2)$ terms will cancel between the
zero-temperature and the temperature-dependent contributions, leaving a
net factor proportional to $\log(\mu^2/T^2)$. After calculating the
contributions from the first and third terms in Eq. (\ref{eq:10}), we
obtain the following result for the temperature-dependent part of $I_0$:
\begin{equation}
  \label{eq:13}
I_0(T) =
  \frac{1}{32\pi^2}\log{\frac{\mu^2}{T^2}} +
\frac{iT}{16\pi \mvk}\log{\frac{k_0+\mvk}{k_0-\mvk}}
+\frac{T}{8\pi i \mvk}\log{\frac{\Gamma(1+K_+/T)}{\Gamma(1+K_-/T)}}.
\end{equation}

Next, consider the $I_1$ integral which can be written as:
\begin{eqnarray}
  \label{eq:14}
I_1=&
\displaystyle{\frac{i\pi}{|\vec k|^3}}
\left(\frac{1}{4\pi}\right)^{\frac{3-\epsilon}{2}}
\displaystyle{\frac{\mu^\epsilon}{\Gamma\left(\frac{3-\epsilon}{2}\right)}}
&\int_{K_{-}}^{K_{+}} dK 
\left[
\frac{k^2}{(4\pi\,K)^2}-\frac{i\,k_0}{2\pi\,K}-1
\right]
\nonumber \\
&{}&\times\int_0^\infty d |\vec q| 
|\vec q|^{1-\epsilon}
\left[1-\displaystyle{\frac{(2\pi\,K)^2}{|\vec q|^2+(2\pi K)^2}}\right]
\left[\displaystyle{\frac 1 2} +
      N\left(\displaystyle{\frac{|\vec q|}{T}}\right)\right].
\end{eqnarray}
Note that the $T=0$ contribution, associated with the factor of $1$ 
in the second square bracket, would apparently lead to a quadratically
divergent integral, which however vanishes in the dimensional
regularization scheme. On the other hand, this factor yields
a leading thermal contribution which is quadratic in the temperature
\begin{equation}
  \label{eq:15}
I_1^{lead}(T)=
\frac{T^2}{24 \mvk^2}\left(1-\frac{k_0}{2\mvk}
\log{\frac{k_0+\mvk}{k_0-\mvk}}\right)
\end{equation}
The $|\vec q|$-integration of the second term
in the second square bracket of Eq. (\ref{eq:14}) is identical
to the one which occurs in $I_0$, so that it gives analogous 
$\log(K)$ contributions which cancel between the $T=0$ and thermal parts.
As we have seen, only such contributions would give rise, after the
$K$-integration, to individual $\log(-k^2)$ terms.
It is possible to evaluate exactly all other temperature-dependent
contributions to $I_1$, in terms of logarithmic functions and
of Riemann's zeta functions with arguments
$(1+ K_\pm/T)$, which are analytic when 
\hbox{$K_\pm\rightarrow 0$ \cite{dealmeida:1992iu}}. Since the complete
expression is rather involved, we indicate here, for simplicity, 
only the logarithmic temperature-dependent contributions to $I_1$:
\begin{equation}
  \label{eq:16}
I_1^{\log}(T) =
  -\frac{1}{192\pi^2}\log{\frac{\mu^2}{T^2}} -
\frac{i T k^2}{64\pi \mvk^3}\log{\frac{k_0+\mvk}{k_0-\mvk}}.
\end{equation}

In a general gauge, $\Pi^\mu_\mu$ and $\Pi_{00}$ will have a similar
behavior (in particular, the leading $T^2$ contribution is gauge
independent). In this case, the thermal contributions to 
\hbox{$\Pi_C=k_\mu \Pi^\mu_0/\mvk$} are non-vanishing, and can be
written as \cite{brandt:1997se}
\begin{equation}
  \label{eq:17}
\Pi_C=\frac{(1-\xi)}{(2\pi)^3\mvk}\int
\frac{d^{3}\vec q}{\vert\vec q\vert}
\left[
\left(\frac{k^2}{k^2+2k\cdot q}+\frac 1 2 \frac{d}{dq_0}
\frac{k\cdot q}{q_0}\right)
\frac{k\cdot q k_0 -k^2 q_0}{k^2+2k\cdot q} N\left(\frac{|q_0|}{T}\right)
+ q\leftrightarrow-q\right]_{q_0=|\vec q|},
\end{equation}
where $\xi$ is the gauge parameter ($\xi=1$ in the Feynman gauge)
and the derivative $d/dq_0$ acts on all terms at its right.
Performing the $|\vec q|$ integration, the terms involving $\log(K)$
factors turn out to be proportional to
\begin{equation}
  \label{eq:18}
  \int_{K_{-}}^{K_+} dK \log K \left[8\pi i K - 3 k_0 -
\frac{k^2 k_0}{16\pi^2K^2}\right].
\end{equation}
However, the coefficient of the $\log(-k^2)$ term, which is obtained
after the $K$-integration is performed, actually vanishes:
\begin{equation}
  \label{eq:19}
\left(2\pi i K_+^2-\frac 3 2 k_0 K_+ -\frac{k^2 k_0}{32\pi^2 K_+}\right)-
\left(2\pi i K_-^2-\frac 3 2 k_0 K_- -\frac{k^2 k_0}{32\pi^2 K_-}\right)
=0
\end{equation}

Thus, the full self-energy of the gluon, which includes the thermal
effects, does not contain $\log(-k^2)$ contributions.
Essentially, these effects replace the zero-temperature
$\log(-k^2/\mu^2)$ term by a $\log(T^2/\mu^2)$ contribution. Although
this correspondence seems plausible, it is not so obvious. For
instance, it would not hold if the thermal contributions would involve
individual terms like $\log(k_0^2/T^2)$, $\log(\mvk^2/k^2)$, etc. To
discard this possibility, it is necessary to show that the $\log(-k^2)$
and $\log(T^2)$ contributions appear in the thermal part only in the
combination $\log(-k^2/T^2)$. Furthermore, in order to explain the 
cancellation of the 
$\log(-k^2)$ terms between the zero-temperature and the
temperature-dependent parts, one must also argue 
\cite{brandt:1997rz} that the $\log(T)$ dependence of the self-energy
is simply related to its ultraviolet behavior at zero-temperature.
Here, these properties of the thermal gluon self-energy have been 
explicitly verified to one-loop order.

\acknowledgements 
We would like to thank CNPq (Brazil) for a grant and
Prof. J. C. Taylor for a helpful correspondence.

\end{document}